\documentclass[10pt,twocolumn,amsmath,amssymb,superscriptaddress,english,aps,pra,longbibliography]{revtex4-2}
\usepackage{graphicx}  
\usepackage{physics}
\usepackage[caption=false]{subfig}
\usepackage{amsmath, amssymb}
\usepackage{bm} 
\usepackage{tabularx}
\usepackage{orcidlink}
\usepackage[export]{adjustbox}

\usepackage{tikz}
\usetikzlibrary{
	matrix, 
	quantikz2,
}

\usepackage{hyperref}
\usepackage{cleveref}
\hypersetup{
colorlinks=true,
urlcolor= black,
citecolor=black,
linkcolor= black,
}

\begin{document}

\title{Demolition measurement protocol for transmon qubits}

\author{Ashutosh Mishra\,\orcidlink{0009-0008-4432-0408}}
\email{a.mishra@fz-juelich.de}
\affiliation{Institute for Quantum Computing Analytics (PGI-12) \\ Forschungszentrum J\"ulich, 52425 J\"ulich, Germany}
\affiliation{Theoretical Physics, Universit\"at des Saarlandes, 66123 Saarbr\"ucken, Germany}

\author{Frank K. Wilhelm\,\orcidlink{0000-0003-1034-8476}}
\affiliation{Institute for Quantum Computing Analytics (PGI-12) \\ Forschungszentrum J\"ulich, 52425 J\"ulich, Germany}
\affiliation{Theoretical Physics, Universit\"at des Saarlandes, 66123 Saarbr\"ucken, Germany}

\author{Shai Machnes}
\affiliation{Qruise GmbH, 66113 Saarbr\"ucken, Germany}
\date{March 31, 2025}

\begin{abstract}
    The process of measuring a qubit and re-initializing it to the ground state practically lead to long qubit idle times between re-runs of experiments on a superconducting quantum computer. Here, we propose a protocol for a \textit{demolition measurement} of a transmon qubit that integrates qubit readout with the reset process to minimize qubit idle time.
    We present a three-staged implementation of this protocol, involving a combined qubit readout and resonator reset scheme that unconditionally resets the resonator at the end of the readout; a leakage removal scheme that can be integrated with the measurement stage; and an unconditional qubit reset. 
    We demonstrate that this protocol could be implemented in $1 \, \mu$s with greater than $95 \, \%$ reset fidelity and a $99 \, \%$ readout fidelity without any hardware overhead beyond those commonly used.
    This provides at least a 50x speed up compared to the passive decay of the qubit, thereby significantly increasing the data-acquisition rate. 
\end{abstract}

\maketitle

\section{Introduction}
    
    Measurement of quantities such as the expectation value of an operator requires executing multiple shots of a circuit on a quantum computer. 
    This kind of measure and re-run scheme is a fundamental part of quantum algorithms like variational quantum eigensolver \cite{tillyVariationalQuantumEigensolver2022}, quantum approximate optimization algorithms \cite{farhiQuantumApproximateOptimization2014}, quantum state and process tomography \cite{poyatosCompleteCharacterizationQuantum1997, chowUniversalQuantumGate2012}, and quantum error correction \cite{fowlerSurfaceCodesPractical2012}.
    In all of these cases, there is a waiting period between measuring the qubit state and running a new circuit. This involves resetting the experiment --- which for a superconducting circuit involves preparing the qubits in the ground state and emptying the readout resonator --- before starting a new set of circuit and measurements. 

    With the increase in qubit relaxation times, passive reset of qubits has been an increasingly inefficient way to bring the qubits to their ground state and necessitates the development of an active reset protocol.     
    There have been efforts to perform active reset of the qubit either conditionally --- by driving the qubit conditional to the measurement outcome \cite{hanActiveResetSuperconducting2023, risteFeedbackControlSolidState2012, salatheLowLatencyDigitalSignal2018}, or unconditionally --- using a sink like a quantum circuit refrigerator (QCR) \cite{aamirThermallyDrivenQuantum2023, hsuQuantumCircuitRefrigerator2021, sevriukFastControlDissipation2019, sevriukInitialExperimentalResults2022}, or the readout resonator \cite{eggerPulsedResetProtocol2018, magnardFastUnconditionalAllMicrowave2018}. 
    Conditional reset protocols are largely affected by measurement inaccuracies and require additional hardware for fast feedback, while QCRs occupy on-chip space. In contrast,
    unconditional reset using the readout resonator incurs no additional hardware overhead and has been demonstrated for both fixed-frequency \cite{eggerPulsedResetProtocol2018, magnardFastUnconditionalAllMicrowave2018, chenFastUnconditionalReset2024a} and tunable qubits \cite{mcewenRemovingLeakageinducedCorrelated2021, zhouRapidUnconditionalParametric2021}.

    In addition to resetting the qubit, resetting the readout resonator and removing the accumulated leakage population from the qubit are essential tasks that must be completed before starting another experiment.
    Processes such as resonator reset \cite{boutinResonatorResetCircuit2017b, bultinkActiveResonatorReset2016, mcclureRapidDrivenReset2016} and qubit leakage reduction \cite{battistelHardwareEfficientLeakageReductionScheme2021, marquesAllMicrowaveLeakageReduction2023} have also been individually demonstrated; however, they have not yet been integrated with qubit readout.
    Resetting each component individually adds significant wait time in experiments, during this the qubit remains in an idle state.
    In this work, we propose a scheme to integrate the reset of a transmon qubit with its readout process, to reduce the waiting time between re-runs of different experiments. 
    We call this a \emph{demolition measurement} of the transmon, which, unlike the conventional projective measurement, resets the state of the qubit to the ground state irrespective of the measurement outcome.
    Considering that measurement is often performed at the end of the circuit and followed by the reset steps, integrating the reset processes with measurement is a pragmatic solution to obtain a compact measurement and reset scheme.
    
    We integrate unconditional qubit reset, using the readout resonator, with a dispersive readout of the qubit. 
    In addition to resetting the computational subspace of the qubit, we propose a method to reset the qubit leakage population that is robust against stray excitations in the resonator. We develop the demolition measurement protocol in stages,  allowing for the integration of different stages with one another, thereby minimizing any qubit idle time in between.
    The scheme we propose here does not require any additional hardware resources beyond those that are already commonly present. 
    We demonstrate that such a scheme could be implemented in around $1 \, \mu$s, hence allowing for an increase in the number of shots that can be performed per second.
    We present the protocol for a fixed-frequency and fixed-coupling architecture, and in future this can be extended and potentially improved for tunable qubits.

    The paper is organized as follows. Section \ref{section: System Description} gives a brief overview of the protocol, the numerical methods used, and a review of the stages in the protocol. In Section \ref{section: Results}, we discuss each stage individually, including the design of the pulses for that stage and how they integrate with the other stages.
    Finally, in Section \ref{Section: Full protocol}, we combine all the stages to construct the demolition measurement protocol.

\section{System Description} \label{section: System Description}

    \begin{figure*}
        \centering
        \includegraphics{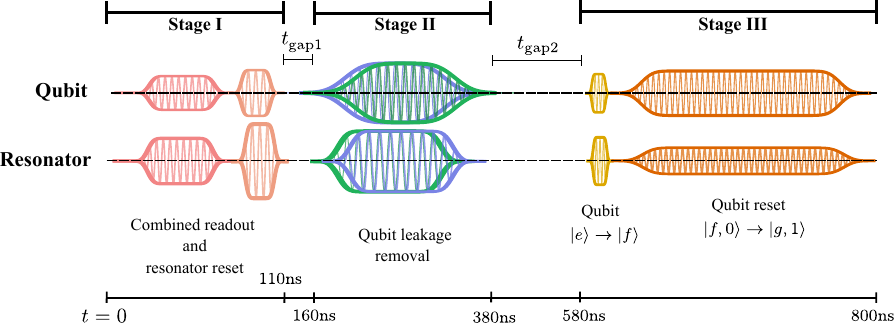}%
        \caption{\textbf{Overview of the protocol:} Pulse scheme for the demolition measurement protocol, demonstrating the three stages - combined readout and resonator reset stage, leakage removal stage, and qubit reset stage. 
        In each stage, the transmon and the resonator are driven by flat-top Gaussian pulses of different frequencies. In stage I, a dispersive measurement of the qubit is performed such that at the end of the measurement the resonator is reset to the ground state. Stage II removes the leakage population of the transmon. Robustness to any residual resonator population is ensured by driving multiple frequency transitions at the same time, shown by the overlap of two different frequency flat-top Gaussian pulses (green and purple pulse). In stage III, qubit reset is performed by utilizing a $\Lambda$ type transition. 
        Appropriate gap times were chosen between the stages (represented by $t_{\text{gap}1}$ and $t_{\text{gap}2}$). The pulse timings represent the ones obtained from the final optimized pulse (not to scale). The pulse shapes, however, are a schematic and do not represent the final shapes.}
        \label{Fig: Full protocol schematics}
    \end{figure*}

    Consider a fixed-frequency transmon capacitively coupled to a readout resonator. The drift Hamiltonian of the transmon-resonator system can be written as,
    \begin{align}
        \begin{split}
            &\hat{H}_\text{drift} = \hat{H}_0 + \hat{H}_c \\
            &\hat{H}_0 = \hbar \omega_r \hat{a}_r^\dagger \hat{a}_r + \hbar \omega_q \hat{a}_q^\dagger \hat{a}_q + \frac{\hbar \delta}{2} \hat{a}_q^\dagger \hat{a}_q^\dagger \hat{a}_q \hat{a}_q \\ 
            &\hat{H}_c  = \hbar g(\hat{a}_q + \hat{a}_q^\dagger)(\hat{a}_r + \hat{a}_r^\dagger) \\ 
        \end{split}
    \end{align}
    where $\hat{H}_0$ represents the bare transmon and resonator Hamiltonian and $\hat{H}_c$ represents the coupling. 
    Here $\hat{a}_q$ and $\hat{a}_r$ represent the annihilation operators of the qubit and resonator, respectively. 
    The transmon anharmonicity is given by $\delta$, which is negative and small (compared to its frequency).
    The coupling is considered to be a fixed dipole-dipole coupling, without rotating wave approximation, with coupling constant $g$. In the dispersive regime, $g$ is designed to be much smaller than the detuning between the qubit and the resonator frequencies $(\Delta = \omega_q - \omega_r)$.

    For controlling the dynamics, both the transmon and the resonator have their separate microwave drive lines. The total Hamiltonian of the system can thus be written as a sum of the drift ($\hat{H}_\text{drift}$) and the drive Hamiltonian ($\hat{H}_d)$
    \begin{align}
        \begin{split} \label{eq1: System Hamiltonian}
            &\hat{H}(t) = \hat{H}_\text{drift} + \hat{H}_d(t) \\
            &\hat{H}_d(t) =  \hbar \Omega_q(t) (\hat{a}_q + \hat{a}_q^\dagger) + \hbar \Omega_r(t) (\hat{a}_r + \hat{a}_r^\dagger)
        \end{split}
    \end{align}
    where $\Omega_q(t)$ and $\Omega_r(t)$ represent the microwave drive fields on the transmon and the resonator, respectively. As the frequencies and coupling of the qubit and resonator are fixed, the only controllable parameters are the microwave drive fields $\Omega_{q/r}(t)$. Thus, the entire protocol for demolition measurement, for fixed-frequency qubits, involves only microwave-activated transitions between the qubit and the resonator. To design pulses for the protocol, we use open-loop quantum optimal control. We design cost functions to achieve suitable goals, and using gradient based optimization techniques tune the parameters of $\Omega_{q/r}(t)$ to minimize the cost.

    For unconditionally resetting the qubit states, and removing the leakage population, we use the resonator as a sink. 
    As the resonator has a low-quality factor (to enable faster readout) \cite{jeffreyFastAccurateState2014}, it decays rapidly as compared to the qubit (in a few $\tau_r \equiv 2\pi/\kappa_r$). This resets the qubit to the ground state irrespective of the measurement outcome and does not require any additional hardware resources.  In this work, we use ``qubit states'' to refer to the lowest two energy levels of the transmon ($\ket{g}, \ket{e}$); ``leakage'' to refer to the second excited level $(\ket{f})$; and ``sink'' to refer to the readout resonator. 

    The demolition measurement protocol involves measuring and unconditionally resetting the qubit, which are inherently dissipative operations. Modeling and optimizing the dissipative dynamics is done by solving the Lindblad master equation,
    \begin{align} \label{Eq: Lindblad master equation}
        \begin{split}
            \dot{\hat{\rho}} = -\frac{i}{\hbar}[\hat{H}, \hat{\rho}] &+ \kappa_q \mathcal{D}[\hat{a}_q]\hat{\rho} + \frac{\gamma_q}{2} \mathcal{D}[\hat{a}_q^\dagger \hat{a}_q]\hat{\rho} + \kappa_q^{\text{th}} \mathcal{D}[\hat{a}_q^\dagger]\hat{\rho} \\ & + \kappa_r \mathcal{D}[\hat{a}_r]\hat{\rho} + \frac{\gamma_r}{2} \mathcal{D}[\hat{a}_r^\dagger \hat{a}_r]\hat{\rho} + \kappa_r^{\text{th}} \mathcal{D}[\hat{a}_r^\dagger]\hat{\rho} 
        \end{split}  
    \end{align}
    where $\mathcal{D}[\hat{a}]\hat{\rho} = \hat{a}\hat{\rho} \hat{a}^\dagger - \frac{1}{2}\{\hat{a}^\dagger \hat{a}, \hat{\rho}\}$ is the Lindblad superoperator, $\kappa_q$ and $\kappa_r$ are the decay rates, and $\gamma_q$ and $\gamma_r$ are the dephasing rates of the qubit and the resonator, respectively.
    Finite temperature effects are included to model re-thermalization of the qubit and the resonator, with  $\kappa_q^{\text{th}}$ and $\kappa_r^{\text{th}}$ representing the thermal excitation rates. As discussed later in Sec. \ref{Subsection: Leakage removal} and Sec. \ref{subsection: Qubit reset}, this can be a source of error for the protocol. To make the model resemble an experimental setting, we have chosen parameter values similar to \cite{ikonenQubitMeasurementMultichannel2019} (as specified in Table \ref{Table: Parameters}).
    While in experiments a Purcell filter is used along with the readout resonator to control the dissipative channels seen by the qubit, in this model we have not explicitly included one. Rather, the transmon decay rate $(\kappa_q)$ is adjusted to match the presence of a Purcell filter.

    In the following subsections, we present a brief outline of the protocol, discuss the numerical methods used, and provide a review of the three stages of the protocol.
    
    \begin{table}
        \begin{center}            
            \begin{tabular}{ || c | c ||}
                \hline
                \textbf{Parameter} & \textbf{Value} \\ [0.5ex]
                \hline \hline
                
                \ Transmon levels \  & \  5 \  \\
                \hline

                \ Transmon frequency ($\omega_q/2\pi$) \  & \  7.86 GHz \  \\
                \hline

                \ Transmon anharmonicity ($\delta$/$2 \pi$) \  & \  -264 MHz \  \\
                \hline

                \ Transmon decay rate ($\kappa_{q}/2\pi$) \  & \  1/(27 $\mu$s) \  \\
                \hline

                \ Transmon dephasing rate ($\gamma_q/2\pi$) \  & \  1/(39 $\mu$s) \  \\
                \hline \hline

                \ Resonator levels \  & \  25 \  \\
                \hline

                \ Resonator frequency ($\omega_r/2\pi$) \  & 6.02 GHz \  \\
                \hline

                \ Resonator decay rate ($\kappa_{r}/2\pi$) $ \equiv \tau_r^{-1}$ \  & \  1/($100$ ns) \  \\
                \hline

                \ Resonator dephasing rate ($\gamma_r/2\pi$) \  & \  1/(50 ns) \  \\
                \hline

                \ Qubit-Resonator coupling strength ($g/2\pi$) \ & \ 130 MHz \ \\
                \hline \hline

                \ Bath temperature \ & \ 50 mK \ \\
                \hline
            \end{tabular}
            \caption{Simulation parameters.}\label{Table: Parameters}
        \end{center}
    \end{table}

    \subsection{Overview of the protocol}\label{subsection: Design of the protocol}

    We propose a protocol that involves three stages, as shown in Fig.~\ref{Fig: Full protocol schematics}.
    In the measurement stage, a fast single-shot dispersive readout is implemented. This involves populating the resonator and measuring the dispersive shift in its frequency.
    Since the consequent stages require an empty resonator to reset the qubit population, we design a combined qubit readout and resonator reset pulse that unconditionally resets the resonator at the end of the measurement.
    
    The second stage involves removing the leakage population from the transmon. This stage is integral to the protocol, as the presence of leakage can lead to errors during qubit reset. We propose a method to remove the leakage population in conjunction with resonator reset by driving multiple transitions involving second-order processes.

    Finally, to reinitialize the qubit to the ground state, an unconditional reset can be performed by driving the qubit excitation to the resonator. The resonator then relaxes to its ground state in a few $\tau_r$, resetting the qubit.

    Combining these three stages measures the state of the qubit and resets it unconditionally to form the demolition measurement protocol.
    The pulses for each stage are constructed such that they are robust to changes in the population of the previous stage.
    We numerically demonstrate that this can be performed in about $1 \, \mu$s, which is more than a 50x speedup compared to the passive transmon decay (comparing to transmons with $T_1 \approx 20-30 \, \mu$s). For transmons with larger $T_1$, the speedup would be even greater.
    
    \subsection{Numerical methods} \label{subsection: numerical approximations}
    
\begin{figure}
        \includegraphics{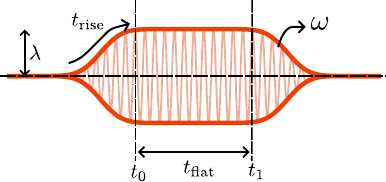}
        \caption{\textbf{Pulse ansatz:} Schematics of a flat-top Gaussian pulse, illustrating the optimizable parameters - rise time ($t_{\text{rise}}$), start ($t_0$) and stop time ($t_1$) of the constant amplitude section, and the frequency ($\omega$) of the pulse. Presence of a small set of optimizable parameters makes this a desirable ansatz for optimization and ensures easy implementation in experimental settings.}
        \label{Fig: flattop gaussian}%
    \end{figure}

    In order to design pulses for the aforementioned stages, we use quantum optimal control techniques. This involves constructing a pulse ansatz and defining suitable goal functions. We use a simple ansatz of smooth pulses - flat-top Gaussian pulse (as shown in Fig.~\ref{Fig: flattop gaussian}), given by,
    \begin{align}
        \begin{split}
            \Omega(t) = \frac{\lambda}{4}\Bigg\{1 + & \text{erf}\Bigg( \frac{t - t_0}{t_{\text{rise}}} \Bigg) \Bigg\} \\ & \Bigg\{1 + \text{erf}\Bigg( \frac{ t_1 - t}{t_{\text{rise}}} \Bigg) \Bigg\} \cos{(\omega t)}
        \end{split}
    \end{align}
    where the pulse amplitude $(\lambda)$, rise time $(t_{\text{rise}})$, start $(t_0)$ and stop time $(t_1)$ of the constant amplitude section, and frequency $(\omega)$ are the optimizable parameters. 
    While for simpler cases having one pulse per subsystem works well for the optimization to converge, this ansatz can be limiting for complex operations.
    To increase the parametrization, we used pulse multiplexing by overlapping multiple pulses per drive line 
    \begin{align}
        \Omega_M(t) = \sum_n \mathcal{E}_n(t) \cos(\omega_n t)
    \end{align}
    where $\Omega_M(t)$ represents the multiplexed pulse and $\mathcal{E}_n(t)$ represents the flat-top Gaussian envelope. 
    Multiplexing different frequency pulses provides the flexibility to explore spaces of complex dynamics while keeping the total number of optimizable parameters small. 
    Additionally, DRAG corrections \cite{motzoiSimplePulsesElimination2009, motzoiImprovingFrequencySelection2013, hyyppaReducingLeakageSingleQubit2024a} were also included to reduce unwanted excitations. 
    
    Since all stages in the protocol are inherently dissipative processes, and operate at a timescale comparable to or longer than the smallest relaxation timescale ($\tau_r$), simulating them requires solving Eq.~\eqref{Eq: Lindblad master equation}. 
    Processes like readout involve driving the resonator close to its resonant frequency, leading to the occupation of high fock states in the resonator. Here we consider 5 levels in the qubit and 25 levels in the resonator for the optimization. The results were further verified by increasing the number of levels.
    
    Traditional optimal control methods require computing the propagator to compute fidelities and gradients. Calculating the propagator for an open-quantum system involves exponentiating the Lindblad superoperator. This scales poorly as the superoperator size grows as $(N^2 \times N^2)$ for a $N$ dimensional system. 
    Thus, we use an ODE-based approach to solve the system dynamics. We use a high-order Runge-Kutta method (Verner's 7th order method \cite{vernerExplicitRungeKuttaMethods1978}), with a fixed time step to ensure that the norm and positivity of the state are preserved during the evolution. Similar methods of using ODE solver for optimization have been used for solving control problems for open systems, with large Hilbert space, to avoid exponentiating superoperators \cite{boutinResonatorResetCircuit2017b, gautierOptimalControlLarge2024a}.
    
    Optimization of the pulses were performed by using the gradient based optimization method L-BFGS \cite{zhuAlgorithm778LBFGSB1997, byrdLimitedMemoryAlgorithm1995}.
    Gradients of the cost function with respect to the optimizable parameters were computed either using automatic differentiation or, in cases when calculating the gradients is expensive, approximated by finite difference.

    \subsection{Qubit readout and resonator reset}
    The dispersive readout of a qubit involves driving the resonator close to its resonant frequency to form a coherent state dependent on the state of the qubit, also known as a pointer state \cite{blaisCircuitQuantumElectrodynamics2021}.
    The phase space trajectory of the pointer state during readout can be used to infer the state of the qubit.
    The overlap between the trajectories of the pointer states across multiple shots, when the qubit is in the ground or excited state, is proportional to the readout error (refer to Appendix~\ref{Appendix: Single-shot trajectories}). 
    Optimizing the readout hence aims at reducing the overlap, which can be achieved by increasing the distance between the pointer state trajectories. We accomplish this by designing a cost function that rewards for larger distance between the pointer states. 

    Using the resonator as a sink, for the subsequent stages, requires an empty resonator.
    Reset of the pointer state by passive decay takes a few $\tau_r$, and adds additional time overhead to the protocol. 
    To shorten the wait time, we introduce an unconditional active reset that clears the resonator in conjunction with qubit readout such that, irrespective of the state of the qubit, the resonator goes to the ground state at the end of the measurement. 
    The entire combined readout and resonator reset pulse can then be integrated to measure the state of the qubit (see Fig. \ref{fig:integrated IQ trajectories}).        
    The combined readout and resonator reset pulse that we present takes about one $\tau_r$. 

\begin{figure*}
            \includegraphics{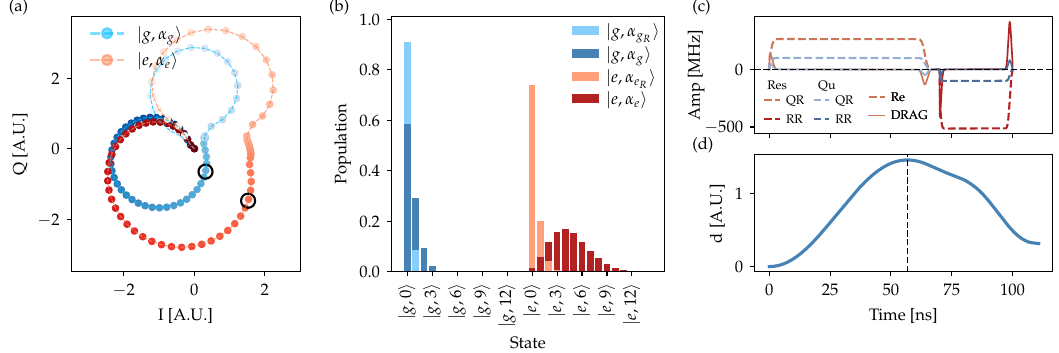}
            
            \caption{\textbf{Stage I : Combined readout and resonator reset:} 
            (a) IQ plane trajectory of the pointer states. The evolution time is represented by the color intensity, darker shade representing the initial time, and lighter shade the final time.
            The pointer state trajectories start and end at the origin, indicating unconditional reset of the resonator. Qubit readout is performed by integrating the trajectory using an optimal weight function. The black circles represent the point of maximum separation of the pointer states.
            (b) Comparison of resonator occupation at the end of resonator reset (lighter color shade) represented by $\ket{g/e, \alpha_{g/e_R}}$, and at the maximal IQ separation (darker color shade) represented by $\ket{g/e, \alpha_{g/e}}$.
            The residual population in the excited states of the resonator, at the end of resonator reset, is dealt with by making the subsequent stages robust to it.
            (c) Optimized pulse envelopes with the DRAG corrections. The dashed lines represent the real part and the solid lines represent the DRAG correction in the optimized pulse. The optimized pulse (for both the qubit and resonator) contains two pulses of different frequencies, the qubit readout pulse (QR) (represented with a lighter shade) and the resonator reset pulse (RR) (represented with a darker shade).
            (d) The IQ plane separation $(d)$ of the pointer states in (a). The black dotted line corresponds to the time of maximum IQ plane separation and represents the time corresponding to black circles in (a).
            }
            \label{Fig: Combined readout and resonator reset}
        \end{figure*}

    Similar active and unconditional resonator reset pulses have been demonstrated, either separately \cite{boutinResonatorResetCircuit2017b, bultinkActiveResonatorReset2016, mcclureRapidDrivenReset2016} or integrated with qubit readout \cite{jergerDispersiveQubitReadout2024a}. While these pulses aim for a resonator population of less than $10^{-2}$ by employing reset pulses longer than 2 to 3$\tau_r$,  we develop a much shorter pulse, leaving a small residual population in the resonator's first and second excited states (see Fig.~\ref{Fig: Combined readout and resonator reset}(b)).
    In the next stage, we build pulses which are robust to this residual population.

    \subsection{Leakage removal}

    While quantum error correcting codes (QEC) can correct errors within the computational subspace, they are ineffective against leakage, which tends to accumulate over QEC cycles \cite{googlequantumaiSuppressingQuantumErrors2023, googlequantumaiQuantumErrorCorrection2024, miaoOvercomingLeakageQuantum2023}.
    Additionally, leakage hinders the qubit reset process. The presence of leakage during qubit reset leads to residual population in the computational subspace, mostly in the $\ket{e}$ state (see Fig.~\ref{Fig: Leakage removal}(e)). 
    While leakage encompasses all the levels higher than $\ket{e}$, leakage errors are predominantly caused by the residual population in the second excited state $(\ket{f})$.
    
    Constructing leakage removal units (LRUs) \cite{battistelHardwareEfficientLeakageReductionScheme2021, marquesAllMicrowaveLeakageReduction2023} to clear the $\ket{f}$ state after each QEC cycle has been an active area of research \cite{campsLeakageMobilitySuperconducting2024, marquesAllMicrowaveLeakageReduction2023, lacroixFastFluxActivatedLeakage2023}. One of the LRUs suitable for fixed-frequency qubits uses the readout resonator as sink for the leakage population \cite{battistelHardwareEfficientLeakageReductionScheme2021, marquesAllMicrowaveLeakageReduction2023}, by driving a transmon-resonator second order transition $\ket{f,0} \rightarrow \ket{g,1}$, which requires an empty resonator.

    Since the resonator retains some residual population at the end of the last stage, we propose a method to remove leakage population from the transmon that is robust to this residual resonator population. 
    This method involves simultaneously driving multiple second order transitions $\ket{f,n} \rightarrow \ket{g, n+1}$ between the transmon and resonator states. 
    
    These transitions can be identified by examining the drive terms of the Hamiltonian under the Schrieffer-Wolff transformation, Eq.~\eqref{eq: reset and leakage Hamiltonian}. It contains terms with $(\hat{a}_r^\dagger \ket{m}\bra{m+2} \, + \, \text{h.c.})$, and represents processes where the qubit loses two photons and the resonator gains one photon and vice versa. 
    So, the small residual population in the resonator after the reset, can be used to simultaneously drive multiple qubit-resonator second-order processes $\ket{f, n} \rightarrow \ket{g, n+1}$. 
    This process transfers the leakage population to the sink in the presence of any resonator population, thereby eliminating the need for long wait times between the two stages or extended reset pulses, while also ensuring robustness against thermal population in the resonator.

    \subsection{Qubit reset}
        An unconditional qubit reset can be performed by swapping the first excited state population of the qubit $(\ket{e})$ with the resonator.
        For fixed-frequency transmons, with fixed coupling, it can be performed by utilizing the $\Lambda$-type transition $(\ket{e,0} \rightarrow \ket{f, 0} \rightarrow \ket{g,1})$ \cite{eggerPulsedResetProtocol2018,battistelHardwareEfficientLeakageReductionScheme2021}, outlined in Fig.~\ref{Fig: Qubit reset}(a).
        After swapping the population, the resonator can decay (in a few $\tau_r$), bringing the entire system to the ground state.
        The effective (driven) coupling for the $\ket{f, 0} \rightarrow \ket{g, 1}$ transition is given by \cite{eggerPulsedResetProtocol2018,zeytinogluMicrowaveinducedAmplitudePhasetunable2015,pechalMicrowaveControlledGenerationShaped2014,battistelHardwareEfficientLeakageReductionScheme2021}
        \begin{align}
            \tilde{g}(t) = \frac{g\delta \Omega(t)}{\sqrt{2} \Delta (\Delta + \delta)}
        \end{align}
        with $\delta$ being the anharmonicity of the qubit, $\Omega(t)$ the drive strength on both the qubit and resonator, and $\Delta = \omega_q - \omega_r$ the detuning between the qubit frequency and the resonator frequency.

\section{Optimizing individual stages} \label{section: Results}

    In this section, we discuss the three stages of the protocol individually and demonstrate how each stage can be optimized to integrate with subsequent stages.

    \subsection{Stage I: Combined Qubit Readout and Resonator Reset} \label{subsection: Dispersive readout}

       The first stage involves performing a combined qubit readout and resonator reset.
       Optimization of the readout pulse was performed by designing pulses such that there is a larger separation between the pointer state trajectories, leading to a smaller overlap and hence a decrease in readout errors. 
       Both the qubit and the resonator were driven with flat-top Gaussian pulses, including tunable DRAG corrections. The qubit drive was added to achieve a faster pointer state separation \cite{gautierOptimalControlLarge2024a, ikonenQubitMeasurementMultichannel2019, touzardGatedConditionalDisplacement2019}.
       The optimized pulse shapes are shown in Fig.~\ref{Fig: Combined readout and resonator reset}(c) and the optimized frequencies are shown in Fig.~\ref{Fig: pulse frequencies}(b).       
        A smoothly varying cost function was used such that minimizing the cost maximizes the distance $d$, 
        \begin{align} \label{Eq: Readout cost, IQ plane distance}
                C_\text{Readout} = 1 - \Phi_{\text{readout}}(\hat{\rho}_g, \hat{\rho}_e) = \min_{t \in [0, t_{\text{f}}]}\qty{e^{-d(t)/d_0}}
        \end{align}
        where $d(t) = \left\Vert \alpha_g(t) - \alpha_e(t) \right\Vert_2$, and $\alpha_g(t)$ $\qty(\alpha_e(t))$ are the expectation values of the annihilation operator of the resonator when the qubit is in ground (excited) state at time $t$, and $t_{\text{f}}$ represents the final time. The parameter $d_0$ is a hyperparameter for the optimization that can be tuned to vary the slope of the cost function.
        And, $\Phi_\text{readout}(\hat{\rho}_g, \hat{\rho}_e)$ represents the readout objective function that we wish to maximize.
        The IQ plane trajectories were computed by computing the expectation values
        \begin{align}
            \begin{split}
                & I = \frac{1}{\sqrt{2}} \langle \hat{a}_r + \hat{a}_r^\dagger \rangle =  \text{Re}\Big( \text{Tr}\{ \hat{\rho} \, \hat{a}_r \} \Big) \\
                & Q = \frac{1}{\sqrt{2}i} \langle \hat{a}_r - \hat{a}_r^\dagger \rangle = \text{Im}\Big( \text{Tr}\{ \hat{\rho} \, \hat{a}_r \} \Big)
            \end{split}
        \end{align}
        and the states $(\hat{\rho})$ were obtained by solving Eq.~\eqref{Eq: Lindblad master equation}, with initializing the qubit in ground and excited state. Experimentally, these are computed by measuring the quadratures of the reflected/transmitted field.
        
        For designing the active resonator reset pulse, reset of the resonator is optimized concurrently with readout optimization, using a multi-goal cost function. 
        It is advantageous to perform a combined optimization as this allows for a fast resonator reset while simultaneously improving qubit state discrimination.
        For the combined optimization, a cost function was used that rewards for a larger IQ distance in between the pulse and penalizes for a higher average resonator population at the end of the pulse,
        \begin{align}
            \begin{split}
                & C_\text{I} = 1 - \Big( \Phi_{\text{readout}}(\hat{\rho}_g, \hat{\rho}_e) \times \Phi_{\text{reset}}(\hat{\rho}_g^f, \hat{\rho}_e^f) \Big) \\
                & \Phi_{\text{readout}}(\hat{\rho}_g, \hat{\rho}_e) = \max_{t\in [0,t_{\text{f}}]}\qty( 1 - e^{-d(t)/d_0} ) \\
                & \Phi_{\text{reset}}(\hat{\rho}_g^f, \hat{\rho}_e^f) =  w_g F(\hat{\rho}_g^f , \hat{\sigma}_g) + w_e F(\hat{\rho}_e^f , \hat{\sigma}_e) \\
                & F(\hat{\rho}, \hat{\sigma}) = \text{Tr}\big( \sqrt{\sqrt{\hat{\rho}} \hat{\sigma} \sqrt{\hat{\rho}}} \big)
            \end{split}
        \end{align}
        where $\Phi_{\text{reset}}(\hat{\rho}_g^f, \hat{\rho}_e^f)$ is the resonator reset objective function representing the average resonator reset fidelity at the end of the pulse ($\hat{\rho}_g^f$ and $\hat{\rho}_e^f$ are the final states obtained by starting from the ground and excited state of qubit respectively), and $F(\hat{\rho}, \hat{\sigma})$ is the overlap between the simulated and target density matrices. $w_g$ and $w_e$ are the normalized weights for the reset process. 
        As discussed in Sec.~\ref{Subsection: Leakage removal}, $w_g$ is considered to be greater than $w_e$ to reduce leakage injection to the qubit.
        Here, $\hat{\sigma}_g = \ket{g, 0}\bra{g, 0}$ and $\hat{\sigma}_e = \ket{e, 0}\bra{e, 0}$ are the target states for resonator reset. This imposes a penalty for any population in the resonator while ensuring that the qubit state remains unaltered.
        A product of the two objective functions guarantees that both conditions of readout and resonator reset are fulfilled simultaneously.
        
        Fig.~\ref{Fig: Combined readout and resonator reset}(a) demonstrates the evolution of the pointer states for the optimized combined readout and resonator reset process. Both pointer states start and end at the origin, while showing a separation in between (see Fig.~\ref{Fig: Combined readout and resonator reset}(d)). This shows an unconditional reset of the resonator irrespective of the initial state of the qubit.
        Fig.~\ref{Fig: Combined readout and resonator reset}(b) compares the resonator population distribution at the time of maximum IQ distance, and at the end of the pulse. The resonator reaches the ground state with a $91\,\%$ fidelity when the qubit is in the ground state, and $74\, \%$ fidelity with the qubit in the excited state. 
        While a longer resonator reset pulse can improve reset fidelity, we demonstrate that a shorter reset pulse, which leaves some residual population in the resonator,  can also be effective by designing the leakage removal stage to be robust against this residual population. This approach shortens the wait time between the two stages. The entire process of readout and resonator reset takes about $100$~ns which corresponds to one $\tau_r$. Finally, we estimate the readout fidelity of the optimized pulse to be $99.8 \%$ by simulating single-shot trajectories, see Appendix~\ref{Appendix: Single-shot trajectories}.

    \subsection{Stage II: Leakage removal}\label{Subsection: Leakage removal}

        \begin{figure*}
            \centering 
            
            \includegraphics{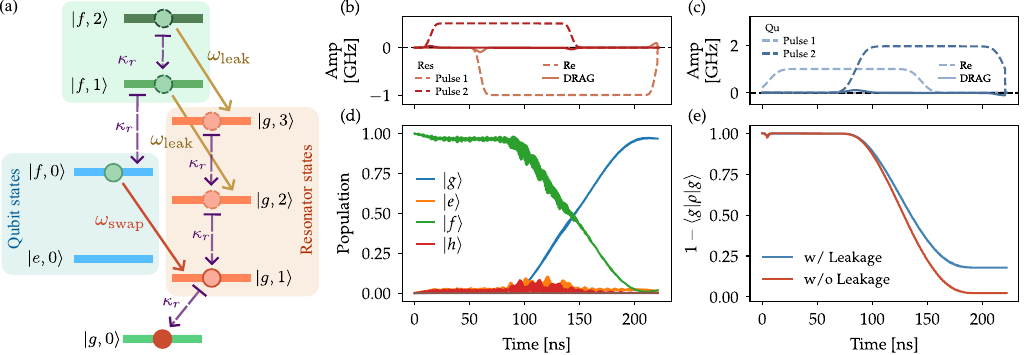}
            
            \caption{\textbf{Stage II: Leakage removal:} (a) Schematics demonstrating the protocol for leakage removal stage. The blue shaded region represents the qubit transitions (with the resonator in the ground state) and orange represents the resonator transitions (with the qubit in the ground state). The green shaded regions represent an excitation of both the qubit and the resonator states. To clear the leakage population, multiple second-order processes $\ket{f, n} \rightarrow \ket{g, n+1}$ (represented by $\omega_{\text{leak}}$ and $\omega_{\text{swap}}$ ) are driven to transfer the qubit population to the resonator. The initial state can be a superposition of $\ket{f, 0}, \ket{f, 1}, \ket{f, 2}$, hence integrating this stage with the measurement stage.
            The resonator states also decay during this process, by rate $\kappa_r$.
            (b), (c) Optimized resonator and qubit drive pulses, respectively. The lighter and darker shade of pulses represent the envelopes of two different frequency pulses overlapped to obtain the final pulse. The dashed line represents the real part and the solid lines represent the DRAG corrections.
            (d) Optimized pulse dynamics on the qubit subspace, starting from the $\ket{f}$ state. The resonator states have been traced out to give a clear demonstration of the qubit dynamics. (e) Comparison of qubit reset in the presence and absence of leakage in the initial state. The presence of leakage leads to an improper reset of the qubit, hence motivating the need for the leakage removal stage.}
            \label{Fig: Leakage removal}
        \end{figure*}
    
        For leakage removal, we simultaneously drive multiple second order transitions $\ket{f, n} \rightarrow \ket{g, n+1}$ between the transmon and the resonator states.
        Fig.~\ref{Fig: Leakage removal}(a) illustrates the leakage removal process, starting with the superposition of the states $\ket{f, 0}, \ket{f, 1}$ and $\ket{f, 2}$ as the initial state. This choice is based on the fact that the residual population from the previous stage includes only the first two excited states of the resonator (see Fig.~\ref{Fig: Combined readout and resonator reset}(b)).
        The second-order processes described above takes these to $\ket{g, 1}, \ket{g, 2}$ and, $\ket{g, 3}$ respectively. 
        These three transitions are driven simultaneously, resetting the $\ket{f}$ state of the transmon to the ground state. The resonator can finally decay (in a few $\tau_r$) to the ground state, removing the leakage population.
        
        Pulses for this process were designed by optimizing the state-overlap fidelity in conjunction with a cost function that quantifies the population of the qubit outside the computational subspace,
        \begin{align}
            & C_\text{II} = 1 - (\bra{g}\hat{\rho}_Q\ket{g} + \bra{e}\hat{\rho}_Q\ket{e})
        \end{align}
        with $\hat{\rho}_Q = \trace_R(\hat{\rho})$ representing the qubit states after tracing out the resonator states.
        Robustness of this pulse to changes in initial qubit and resonator state was ensured by performing an ensemble optimization, by starting with four different initial states, given by
        \begin{align}
        \begin{split}
            \hat{\rho}_1 &= \hat{\rho}_{\mathbf{I}}(T) \\ 
            \hat{\rho}_2 & = \ket{\psi_2}\bra{\psi_2}, \ \ket{\psi_2} = \qty{\alpha_1\ket{f,0} + \alpha_2\ket{f, 1} + \alpha_3\ket{f, 2}} \\
            \hat{\rho}_3 &= \ket{\psi_3}\bra{\psi_3}, \ \ket{\psi_3} = \qty{\beta_1\ket{g,0} + \beta_2\ket{e, 0} + \beta_3\ket{f, 0}} \\
            \hat{\rho}_4 &= \ket{\psi_4}\bra{\psi_4}, \ \ket{\psi_4} = \qty{\gamma_1\ket{g,0} + \gamma_2\ket{g, 1} + \gamma_3\ket{g, 2}} \\
        \end{split}
        \end{align}
        where $\hat{\rho}_{\mathbf{I}}(T)$ represents the mixed state obtained at the end of Stage I. The parameters $\alpha_i$, $\beta_i$, and $\gamma_i$ represent weights for the states in the superposition, and are chosen to resemble errors in an experimental setting. 
        The state $\hat{\rho}_2$ represents a fully leaked qubit with the resonator in a superposition state, while $\hat{\rho}_3$ represents a small leakage in the qubit outside the computational subspace (with $\beta_3 < \beta_1, \beta_2$).
        The state $\hat{\rho}_4$ represents the qubit in the ground state with some residual population in the resonator, and is included to penalize for leakage injection into the qubit. 
        An average of the fidelities obtained from the four states was then used as a metric for optimization.
        
        For the optimization to reach fidelities greater than $95\,\%$, we employed pulse multiplexing by overlapping two flat-top Gaussian pulses with different frequencies. The optimized pulse envelopes for the qubit and the resonator are illustrated in Fig.~\ref{Fig: Leakage removal}(b), (c), depicting envelopes of the two different frequency pulses. Fig.~\ref{Fig: pulse frequencies}(c) illustrates the optimized frequency spectrum of the these two pulses.
        The qubit dynamics (after tracing out the resonator states) is shown in Fig.~\ref{Fig: Leakage removal}(d). It demonstrates the leakage removal process with a $98\,\%$ fidelity.

        The leakage removal (and the qubit reset) process is susceptible to the presence of residual resonator population when the qubit is in the ground state. This can lead to injection of population back into the $\ket{f}$ state of the qubit, by driving the $\ket{g, n} \rightarrow \ket{f, n-1}$ transition. This issue also plagues the state-of-the-art leakage removal processes \cite{battistelHardwareEfficientLeakageReductionScheme2021, marquesAllMicrowaveLeakageReduction2023}, and strictly requires an empty resonator when the qubit is in the ground state. 
        We partially mitigate this by a small wait time between Stage~I and Stage~II.
        
        Moreover, a higher weight was assigned to the resonator reset optimization (as discussed in Sec. \ref{subsection: Dispersive readout}) when the qubit is initialized in the ground state, ensuring that the resonator is properly reset under these conditions. Additionally, the state $\hat{\rho}_4$ was incorporated into the ensemble optimization for leakage removal as a penalizing term. Together, these measures ensure a readout with resonator reset and leakage removal, reintroducing minimal leakage into the qubit. 
        
    \subsection{Stage III: Qubit reset} \label{subsection: Qubit reset}

        \begin{figure*}
            \centering
            \includegraphics{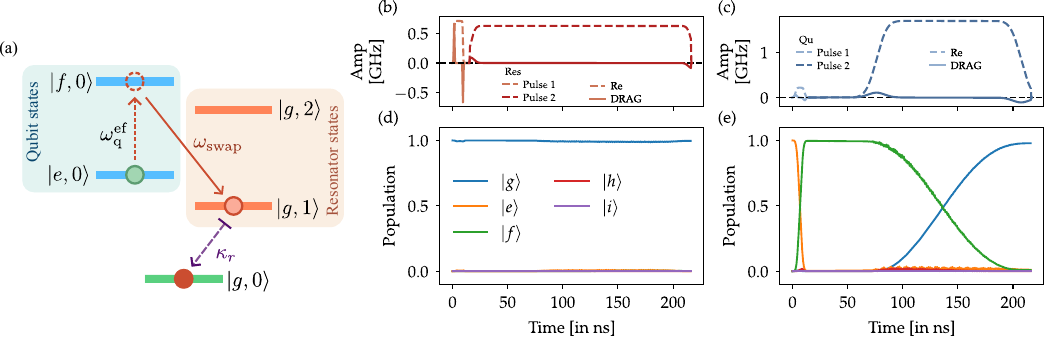}
            \caption{\textbf{Stage III: Qubit reset:} (a) Schematics demonstrating the qubit reset protocol using the $\Lambda$-type transition $(\ket{e,0} \rightarrow \ket{f,0} \rightarrow \ket{g, 1})$. Blue shaded regions demonstrate the qubit transition, and orange shaded ones are for the resonator. A second-order process connects the qubit and resonator transition (represented by $\omega_{\text{swap}}$). The resonator population can finally decay (with rate $\kappa_r$) bringing the entire system to the ground state. (b), (c) Optimized control pulses for the resonator and the qubit, respectively, with the real part demonstrated by the dashed lines and the DRAG corrections by the solid lines. The lighter and darker shade of pulses represent pulses corresponding to the $\ket{e, 0} \rightarrow \ket{f,0} $ and $\ket{f, 0} \rightarrow \ket{g,1}$ transition, respectively.  (d), (e) Qubit dynamics (after tracing out the resonator states) under the optimized pulse for qubit starting in the ground and the excited state, respectively. The qubit stays in the ground state if it starts in the ground state, but reset to ground state if it starts in the excited state, performing an unconditional reset.}
            \label{Fig: Qubit reset}
        \end{figure*}
        
        For the third stage, we design the qubit reset pulse by optimizing for the $\Lambda$-type transition $(\ket{e,0} \rightarrow \ket{f, 0} \rightarrow \ket{g,1})$.
        We use a reset cost function that penalizes for any population outside the ground state of the qubit,
        \begin{align}
            C_\text{Reset} = 1 - \bra{g}\hat{\rho}_Q\ket{g}
        \end{align}
        with $\hat{\rho}_Q = \trace_R(\hat{\rho})$, the qubit state after tracing out the resonator. Moreover, similar to the previous stage, an ensemble optimization was performed using four initial (and target) states by including small leakage and resonator population to make the pulses robust to changes in the initial state. This makes it easier to combine it with the previous stages. 
    
        A combination of two different frequency pulses is used for this stage, corresponding to the two transitions. Fig.~\ref{Fig: Qubit reset}(b) and (c) represent the optimized pulses with the lighter shade pulse representing the $\ket{e,0} \rightarrow \ket{f,0}$ transition and the darker shade pulse representing the $\ket{f,0} \rightarrow \ket{g,1}$ transition, and Fig.~\ref{Fig: pulse frequencies}(c) demonstrates the frequency spectrum of the pulse.
        Fig.~\ref{Fig: Qubit reset}(d) and (e) demonstrate the qubit populations (after tracing out the resonator states) under the optimized reset pulse. The qubit is preserved in the ground state with a fidelity of $99.7\,\%$ when the qubit starts in the ground state, and reset to the ground state with $97.9\,\%$ fidelity when starting the qubit in the excited state.
        With this, we perform an unconditional reset on the qubit after measuring the qubit state. After the qubit reset, the excited state of the resonator can decay to the ground state in a few $\tau_r$, thus resetting both the qubit and the resonator to the ground state.

        Similar to the previous stage, the presence of thermal population in the resonator can induce leakage in the qubit, if the qubit starts in the ground state. While such a state was considered in the ensemble optimization to penalize for any induced leakage, re-thermalization of the resonator during the qubit reset process still contributes to some errors in qubit reset.

\section{Demolition measurement}\label{Section: Full protocol}

    \begin{figure}
        \centering
        \includegraphics{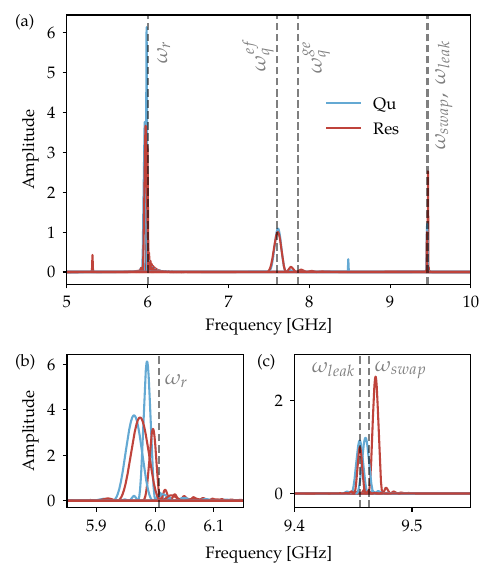}
        \caption{\textbf{Optimized pulse frequencies:} (a) Frequency spectrum after combining the optimized pulses from all three stages. The labelled frequencies represent the eigen-frequencies of the Hamiltonian under a constant drive of 2~GHz on the qubit and 0.5~GHz on the resonator. $\omega_q^{\text{ge}}$ and $\omega_q^{\text{ef}}$ represent the $\ket{g}$ $\rightarrow$ $\ket{e}$ and, $\ket{e}$ $\rightarrow$ $\ket{f}$ transition frequencies, respectively. $\omega_r$ represents the dressed resonator frequency. $\omega_{\text{swap}}$ and $\omega_{\text{leak}}$ represents the $\ket{f, 0}$ $\rightarrow$ $\ket{g, 1}$ and $\ket{f, 1}$ $\rightarrow$ $\ket{g, 2}$ transition frequency, respectively.
        (b), (c) represent the zoomed in spectrum centered around 6.0~GHz and 9.475~GHz respectively. (b) corresponds to the readout and resonator reset pulse, and (c) corresponds to the pulses for leakage removal and qubit reset stages.}
        \label{Fig: pulse frequencies}
    \end{figure}

    \begin{figure}
        \centering
        \includegraphics{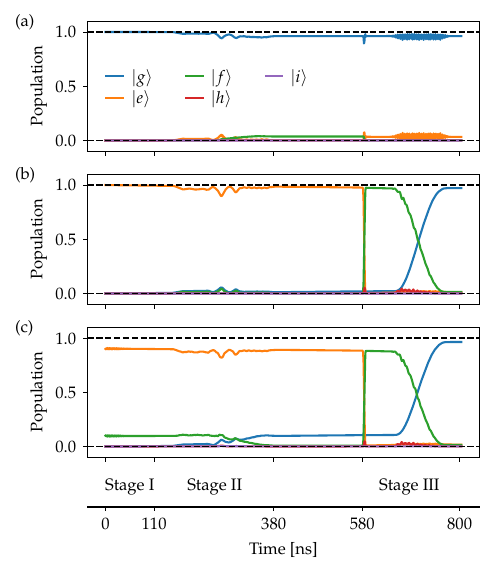}
        \caption{\textbf{Combining the three stages:} (a), (b), and (c) represent the qubit dynamics (after tracing out the resonator states) with the qubit starting from the ground, excited, and excited with $10\,\%$ coherent leakage respectively. In all three cases, at the end of the pulse, the qubit returns to the ground state with fidelity $96.2\,\%$, $97.1\,\%$, $96.3\,\%$ respectively. In the presence of leakage, the leakage population is reset to the ground state before the qubit reset stage.}
        \label{Fig: full protocol}
    \end{figure}

    To construct the demolition measurement protocol, we combine the three stages while tuning the gaps between the stages. 
    Since the leakage removal stage requires the resonator to be empty when the qubit is in the ground state, a 50~ns $(0.5\tau_r)$ gap is added between the first and the second stage so that the small residual population can decay. 
    This clears the resonator when the qubit is in the ground state, preventing leakage injection into the qubit.
    While the resonator reset process was optimized keeping this process in mind, the extra gap time increases the robustness of the pulse.
    
    Moreover, the qubit reset stage is also susceptible to injection of leakage to the qubit in the presence of residual population in the resonator. Hence, after the end of the leakage removal stage, we add an extra 200~ns $(2\tau_r)$ gap time, such that the excited states of the resonator can decay. 
    These can be seen in Fig.~\ref{Fig: Full protocol schematics} as $t_\text{gap1}$ and $t_\text{gap2}$.
    As the various stages involve pulses of different frequencies, we show the frequency spectrum of the entire pulse in Fig.~\ref{Fig: pulse frequencies}. 
    We compare the frequency spectrum with the eigenfrequencies of the driven Hamiltonian under a constant drive of amplitude 2~GHz on the qubit and 0.5~GHz on the resonator.
    It can be seen that the optimized frequencies are slightly shifted from the resonance frequencies of the transitions. This can partly be attributed to different stark shift due to the varying amplitude of the pulse.
    For readout (Fig.~\ref{Fig: pulse frequencies}(b)), both the qubit and the resonator are driven at a frequency lower than $\omega_r$. For the swap and leakage removal (Fig.~\ref{Fig: pulse frequencies}(c)), the qubit is driven at the frequency of $\ket{f,1} \rightarrow \ket{g,2}$ transition, while the resonator is driven at a slightly higher frequency. Additionally, small peaks in the spectrum, corresponding to the leakage removal pulse in Stage II, where the resonator is driven about 5.5~GHz, and the qubit is driven around 8.5~GHz, could not be identified with any transition frequency. These frequency shifts require further investigation.
    
    In Fig.~\ref{Fig: full protocol}(c) and (d) the dynamics of the qubit states (after tracing out the resonator states) under the demolition measurement protocol are shown by initializing the qubit in the ground and the excited states respectively. 
    In both cases, the qubit is reset to the ground state at the end of the pulse with fidelity $96.2\,\%$ and $97.1\,\%$. To test the robustness of the pulse to some initial leakage population in the qubit, we simulated the dynamics by initializing the qubit in the excited state with $10\,\%$ coherent leakage population in the second excited state. Fig.~\ref{Fig: full protocol}(e) shows the evolution of this state. Here, the second stage resets the leakage population to the ground state, and the qubit is reset to the ground state at the end of the pulse with $96.3\,\%$ fidelity. The readout fidelity at the end of Stage I was further estimated to be 99.8 \%, by simulating single-shot trajectories (see Appendix \ref{Appendix: Single-shot trajectories}). 

    This demonstrates that the entire process of measuring the qubit, removing any accumulated leakage and resetting the qubit and resonator to the ground state can be performed under $1\,\mu$s. While we demonstrate greater than $ 95\,\%$ fidelity of reset, the remaining infidelity mainly stems from re-thermalization of the resonator during the qubit reset stage.

\section{Conclusion}\label{section: conclusion}
    In this work, we presented a protocol that integrates qubit reset with the measurement of the qubit state. We refer to this as the demolition measurement protocol, and demonstrated that, for fixed-frequency transmons, in about $1 \,\mu$s one can measure the state of the qubit and reset the entire setup (both the qubit and the readout resonator) to the ground state.  
    This reduces the time overhead between successive runs of a quantum circuit from over $50-100 \, \mu$s to about $1 \, \mu$s, significantly increasing the data acquisition rate in experiments. 
    As our scheme does not require any additional hardware, either in the control stack or on-chip, it does not affect the scalability of the quantum processor. 
    
    Our approach utilizes a three-stage protocol that includes the combined qubit readout and resonator reset stage, the leakage removal stage, and the qubit reset stage. The readout resonator serves both to measure the state of the qubit and as a sink to unconditionally remove the qubit excitations. In contrast to executing the readout and reset processes individually, we design each stage to integrate with the previous one.
    
    In comparison to the leakage reduction schemes proposed in previous works \cite{battistelHardwareEfficientLeakageReductionScheme2021,
    marquesAllMicrowaveLeakageReduction2023}, our protocol integrates leakage removal with measurement. This approach can also be used to prevent the accumulation of leakage in quantum algorithms and quantum error correction by clearing the leakage population after each readout without introducing significant time overhead. 

    The pulses for the three stages were designed using quantum optimal control tools, by formulating suitable cost functions and ensuring robustness of the pulse through ensemble optimization.
    The selection of a simple pulse ansatz was motivated by its ease of implementation in experimental settings. While an ansatz with more parameters, like the piece-wise constant ansatz, can yield higher fidelities, it is often challenging to implement experimentally and is more sensitive to changes in model parameters as compared to a simple pulse shape. 

    Although we have demonstrated the demolition measurement exclusively for fixed-frequency qubits, with fixed couplings, this approach can also be extended to tunable qubits. Owing to the extra degree of freedom, from the tunability of the qubit frequency, faster methods could be designed.
    Furthermore, this speedup could potentially mitigate the issue of leakage injection into the qubit resulting from the re-thermalization of the resonator.
    In the future, the concept of shelving prior to readout \cite{chenTransmonQubitReadout2023} could also be explored to further compact the protocol.

\begin{acknowledgments}
    We thank our colleagues, particularly Alessandro Ciani and Nicolas Wittler for discussions, comments, and review of the manuscript. 
    This work is supported by funding from the German Federal Ministry of Education and Research via the funding program quantum technologies – from basic research to the market – under contract numbers 13N15680 ``GeQCoS'' and HVF-0096 ``Qruise HVF''.    
\end{acknowledgments}

\appendix
\section{Single-shot trajectories} \label{Appendix: Single-shot trajectories}
    To estimate the readout fidelity, we emulate the single shot readout process by using quantum trajectories obtained from solving the stochastic master equation (SME) \cite{jacobsStraightforwardIntroductionContinuous2006} 
    \begin{align}
        \begin{split}
            d\hat{\rho} = &-\frac{i}{\hbar}[\hat{H}, \hat{\rho}] +  \sum_i \mathcal{D}[\hat{L}_i]\hat{\rho} \, dt 
            \\ &+ \sum_n \big( \mathcal{D}[\hat{M}_n]\hat{\rho} \, dt + \sqrt{\eta_n} \mathcal{H}[\hat{M}_n] \hat{\rho} \, dW \big)
        \end{split}
    \end{align}
    where $\{\hat{L}_i\}$ represent the jump operators which describe the dissipative processes, with the Lindblad superoperators given by
    \begin{align}
        \mathcal{D}[\hat{L}_i]\hat{\rho} \coloneqq \hat{L}_i \hat{\rho} \hat{L}_i^\dagger - \frac{1}{2} \big( \hat{L}_i^\dagger \hat{L}_i \hat{\rho} + \hat{\rho} \hat{L}_i^\dagger \hat{L}_i \big)
    \end{align}
    and $\{\hat{M}_n\}$ represent the measurement operators, with the measurement superoperator given by
    \begin{align}
        \mathcal{H}[\hat{M}_n]\hat{\rho} \coloneqq \hat{M}_n \hat{\rho} + \hat{\rho} \hat{M}_n^\dagger - \langle \hat{M}_n + \hat{M}_n^\dagger \rangle \hat{\rho}
    \end{align}
    $\{\eta_n\}$ represents the efficiency of the measurement channels. 

    These trajectories can then be used to measure the qubit state, by integrating the IQ plane signal using the optimal weight functions \cite{bultinkGeneralMethodExtracting2018}, given by
    \begin{align}
        \begin{split}
            w_I(t) &= \langle I_{\ket{e}}(t) - I_{\ket{g}}(t)  \rangle \\
            w_Q(t) &= \langle Q_{\ket{e}}(t) - Q_{\ket{g}}(t)  \rangle
        \end{split}
    \end{align}
    where averaging is performed over trajectories.
    The integrated IQ signal $(S(t))$ can then be written as
    \begin{align}
        \begin{split}
            S_{\ket{g}}(t) &= \int_0^t w_I(t')I_{\ket{g}}(t') + w_Q(t')Q_{\ket{g}}(t') \, dt' \\
            S_{\ket{e}}(t) &= \int_0^t w_I(t')I_{\ket{e}}(t') + w_Q(t')Q_{\ket{e}}(t') \, dt'
        \end{split}
    \end{align}.
    
    \begin{figure}
        \centering
        \includegraphics{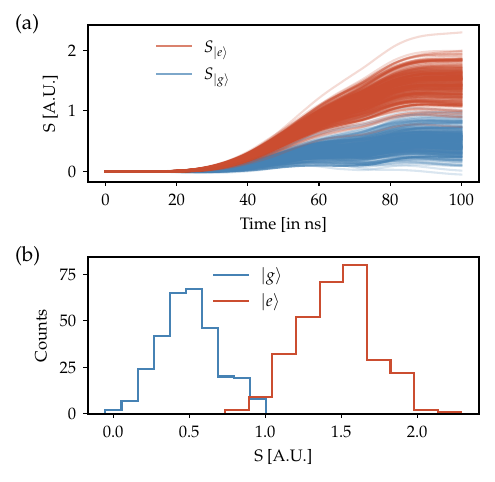}
        \caption{\textbf{Estimating readout fidelity} (a) Integrated IQ signal using the optimal weight function of 300 shots of SME each for the qubit starting in ground and excited state. (b) Histogram of the spread of integrated IQ signal at the end of the readout pulse. The region of overlap represents the error in readout.}
        \label{fig:integrated IQ trajectories}
    \end{figure}

    For the simulation, we consider a measurement efficiency $\eta = 1.0$.
    Fig.~\ref{fig:integrated IQ trajectories} demonstrates integrated IQ signal for 600 simulated trajectories and the distribution at the end of the readout. A clear separation between the ground and the excited states can be seen. The overlap in Fig.~\ref{fig:integrated IQ trajectories}(b) accounts for the readout error.  
    The signal-to-noise ratio (SNR) can be computed as \cite{bultinkGeneralMethodExtracting2018} 
    
    \begin{align}
        \text{SNR} = \frac{\big\lvert \langle S_{\ket{e}} - S_{\ket{g}} \rangle\big\rvert}{\sqrt{\langle S^2 \rangle - \langle S\rangle^2}}
    \end{align}
    where the noise is computed as the average over the deviation obtained by the two states.
    And the readout fidelity can be estimated using the SNR as \cite{blaisCircuitQuantumElectrodynamics2021, gambettaProtocolsOptimalReadout2007}
    \begin{align}\label{eq: readout fidelity}
        F_r = 1 - \text{erfc}\qty(\frac{\text{SNR}}{2})
    \end{align}
    where `erfc' is the complementary error function. The computed SNR using the 600 shots of SME is $4.45$ and readout fidelity estimated using Eq.~\eqref{eq: readout fidelity} is $99.8\%$.

\section{Qubit-Resonator transition} \label{Appendix: Leakage removal}
    Starting from the Hamiltonian in Eq. \eqref{eq1: System Hamiltonian} and following \cite{battistelHardwareEfficientLeakageReductionScheme2021} to calculate the effective driven-coupling strength between the qubit and the resonator levels, we can perform a Schrieffer-Wolff transformation to the system Hamiltonian to the first order in the perturbation parameter $\frac{g}{\omega_q - \omega_r}$. 
    Here the total Hamiltonian includes only the qubit drive.
    As shown by \cite{battistelHardwareEfficientLeakageReductionScheme2021}, a transformed Hamiltonian $(\hat{H}^S)$ can be obtained
    \begin{align}
        \hat{H}^S = \hat{H}_\text{drift}^S + \hat{H}_{d1}^S + \hat{H}_{d2}^S
    \end{align}
    where
    \begin{align}
        \begin{split}
            \hat{H}_\text{drift}^S = & \Big( \zeta_r - \sum_{m=0}^{\infty} \frac{g^2 \Delta_{-1}}{\Delta_m \Delta_{m-1}} \ket{m}\bra{m} \Big) \hat{a}_r^\dagger \hat{a}_r  \\& + \sum_{m=1}^{\infty} \Big( m\zeta_q + \frac{\delta}{2} m(m-1) + \frac{g^2 m}{\Delta_{m-1}} \Big)\ket{m}\bra{m}
        \end{split}
    \end{align}
    is the diagonal part of the Hamiltonian and contains dispersive shifts. The frequencies $\zeta_r = \omega_r - \omega_d$ and $\zeta_q = \omega_q - \omega_d$ represent the detuning from the drive frequency $(\omega_d)$, and $\Delta_{m} = \Delta + \delta m$.
    Here ${\ket{m}}$ represent the states of the transmon. The term
    \begin{align}
        \hat{H}_{d1}^S = \frac{\Omega e^{i\phi}}{2} \hat{a}_q + \text{h.c.}
    \end{align}
    is the transmon drive in the transformed frame. And
    \begin{align}\label{eq: reset and leakage Hamiltonian}
        \begin{split}
            \hat{H}_{d2}^S = & \frac{\Omega e^{i\phi}}{2} \Big( \hat{a}_r \sum_{m=0}^{\infty} \frac{g \Delta_{-1}}{\Delta_m \Delta_{m-1}} \ket{m}\bra{m} \\& + \hat{a}_r^\dagger \sum_{m=0}^{\infty} \frac{g\delta\sqrt{m+1}\sqrt{m+2}}{\Delta_m \Delta_{m+1}}\ket{m}\bra{m+2} \Big) + \text{h.c.}
        \end{split}
    \end{align}
    contains indirect drive on the resonator due to qubit drive and qubit-resonator coupling. The last term in the Hamiltonian contains terms proportional to $\big(\hat{a}_r^\dagger \ket{m} \bra{m+2} + \hat{a}_r \ket{m+2}\bra{m} \big)$ which can be used to drive the qubit population to the resonator. This is used in both Stage II and Stage III, to remove the leakage population by driving transitions like $\ket{f,n} \rightarrow \ket{g, n+1}$ and to reset the qubit by using the $\ket{f,0} \rightarrow \ket{g,1}$ transition. Although this analysis is performed by only considering a transmon drive, we use an additional drive on the resonator, in the simulations, for more flexibility of optimization.

\end{document}